\documentclass[]{raa}
\usepackage{graphicx,times}
\usepackage{natbib}

\providecommand{\bm}[1]{\mbox{\boldmath$#1$\unboldmath}}

\begin{document}

   \title{Observations of EUV and Soft X-ray Recurring Jets in an Active Region
}

 \volnopage{ {\bf 2009} Vol.\ {\bf 9} No. {\bf XX}, 000--000}
   \setcounter{page}{1}

   \author{Liheng Yang
      \inst{1,2}
   \and Yunchun Jiang
      \inst{1}
   \and Jiayan Yang
      \inst{1}
   \and Yi Bi
      \inst{1}
   \and Ruisheng Zheng
      \inst{1}
   \and Junchao Hong
      \inst{1}
        }

   \institute{National Astronomical Observatories/Yunnan Observatory, Chinese Academy of Sciences,
             Kunming 650011, China; {\it yangliheng@ynao.ac.cn}\\
             \and
             Graduate school of Chinese Academy of Sciences, Beijing 100049, China
   {\small Received [year] [month] [day]; accepted [year] [month] [day] }
}

\abstract{We present simultaneous observations of three recurring
jets in EUV and soft X-ray (SXR), which occurred in an active region
on 2007 June 5. By comparing their morphological and kinematic
characteristics in these two different wavelengths, we found that
EUV and SXR jets had similar locations, directions, sizes and
velocities. We also analyzed their spectral properties by using six
spectral lines from the EUV Imaging Spectrometer (EIS) on board
$\emph{Hinode}$, and found that these jets had temperatures from
0.05 to 2.0 MK and maximum electron densities from
6.6$\times$10$^{9}$ to 3.4$\times$10$^{10}$ cm$^{-3}$. For each jet,
an elongated blue-shifted component and a red-shifted component at
the jet base were simultaneously observed in Fe{\sc xii}
$\lambda$195 and He{\sc ii} $\lambda$256 lines. The three jets had
maximum Doppler velocities from 25 to 121 km s$^{-1}$ in Fe{\sc xii}
$\lambda$195 line and from 115 to 232 km s $^{-1}$ in He{\sc ii}
$\lambda$256 line. They had maximum non-thermal velocities from 98
to 181 km s$^{-1}$ in Fe{\sc xii} $\lambda$195 line and from 196 to
399 km s$^{-1}$ in He{\sc ii} $\lambda$256 line. We also examined
the relationship between averaged Doppler velocities and maximum
ionization temperatures of these three jets, and found that averaged
Doppler velocities decreased with the increase of maximum ionization
temperatures. In the photosphere, magnetic flux emergences and
cancellations continuously took place at the jet base. These
observational results were consistent with the magnetic reconnection
jet model that magnetic reconnection between emerging magnetic flux
and ambient magnetic field occurred in the lower atmosphere.
 }

   \authorrunning{L. H. Yang et al. }
   \titlerunning{EUV and soft X-ray jets}
   \maketitle

\section{Introduction}
\label{sect:intro}

Jets are considered to be ubiquitous phenomena on the solar
surface~\citep{shi07}. They have been extensively observed with
various wavelengthes, such as H$\alpha$~\citep{roy73}, Ca {\sc ii}
H~\citep{shi07,chi08b,nis08}, EUV~\citep{ale99}, and soft X-ray
(SXR,~\citet{shi92}), which reflect their multi-thermal
structures~\citep{chi08b}. X-ray jets were discovered by the Soft
X-ray Telescope (SXT) aboard $\emph{Yohkoh}$~\citep{tsu91}, as
transitory X-ray enhancements with apparent collimated motions
associated with flares (microflares or subflares) in X-ray bright
points (XBPs), emerging flux regions (EFRs), or active regions
(ARs)~\citep{shi92}. Jets from XBPs in ARs tend to appear at the
western edge of preceding sunspots~\citep{shim96} and exhibit a
recurrent nature. The XBP-like structures at the jet bases are
generally rooted in the mixed or satellite magnetic polarities
region~\citep{shim98}. These X-ray jets are typically (1-40)
${\times}$ 10$^{4}$ km in length and have apparent velocities of
10-1000 km s$^{-1}$ and lifetimes of 100s-16000s~\citep{shim96}.

EUV jets are straight or slightly curved ejections observed in the
transition region. They might be bright~\citep{cha99, che08},
dark~\citep{liu04}, or dark-and-bright~\citep{ko05, jia07, nis08}
structures. EUV jets have been extensively studied by many
researchers, but their prevalent properties have not been obtained
due to the large temperature range of the transition region.
Previous observations showed that they might be smaller and shorter
lived than SXR jets~\citep{cha99}. The relationship of jets at
different wavelengths has been studied for a long time, mainly
concentrated on H$\alpha$ and SXR~\citep{rus77,sch94,can96},
H$\alpha$ and EUV~\citep{sch81,cha99,jia07,che08}, and EUV and
SXR~\citep{ale99,kim07,chi08b}. In fact, more studies care about the
former two issues, which disclose the connections of the typical
cool surges and hot jets. However, the associations of EUV and SXR
jets are only studied by several researchers. \citet{ale99} made a
comparison of jets observed at both EUV and SXR. They found that the
geometry, role of hot and cool plasma, ejection speed and blob
characteristics were in qualitative agreement with the magnetic
reconnection model proposed by \citet{yok95,yok96}. However, their
observations were limited in large time cadence and low spatial
resolution of the X-ray images. Following their observations,
\citet{kim07} investigated three AR jets simultaneously observed by
$\emph{Hinode}$ XRT and $\emph{TRACE}$, and they discovered that EUV
and SXR jets had similar projected speeds, lifetimes and sizes.
Recent observations of \citet{chi08b} showed that the EUV jet had
the same location, direction and collimated shape as that of the SXR
jet. However, more examples referred to the relationship between hot
X-ray jets and cooler EUV jets are needed.

Blue-shifted and red-shifted features associated with jet event were
early observed in H$\alpha$~\citep{can96} or H$\beta$~\citep{zha00}
Dopplergrams. These dynamic features were also observed by
spectroscopic instrument, such as the EUV Imaging Spectrometry (EIS)
on board the $\emph{Hinode}$. \citet{kim07} observed an AR jet with
four spectral lines of $\emph{Hinode}$ EIS and found simultaneous
blue-shifted (up to -64 km s$^{-1}$) and red-shifted (up to 20 km
s$^{-1}$) motions at the jet base. Then, \citet{chi08a} reported a
strong blue-shifted component exceeding 150 km s$^{-1}$ associated
with an AR jet and a weak red-shifted component at the jet base. The
EIS instrument can also provide the diagnostics of density and
non-thermal velocity. The recent observations revealed that AR jets
had densities up to 10$^{11}$ cm$^{-3}$~\citep{chi08a}. The
non-thermal velocity of the jet was found to be about 100 km
s$^{-1}$, which was slightly smaller than typical non-thermal speed
of flares~\citep{kim07}. The spectroscopic properties of jets can
provide useful parameters for establishments of the jet models and
need to be further studied.

Previous studies have showed that jets were often associated with
some photospheric magnetic activities at their base, such as
satellite spots~\citep{jia07,che08}, moving magnetic
features~\citep{bro07}, moving magnetic bipoles~\citep{can96}, and
sunspots decay~\citep{che09}. According to~\citet{shim98},
jet-producing areas favored regions of evolving magnetic flux
(increasing or decreasing). They thus suggested that the difference
in the rate of the magnetic flux emergence and photospheric
reconnection led to such observed increase or decrease.
\citet{cha99}, \citet{zha00}, \citet{liu04}, and \citet{jia07}
further found that jets and associated H$\alpha$ surges occurred at
sites where the pre-existing magnetic flux was canceled with the
newly emerging flux of opposite polarity, and they inclined to
believe that flux cancellation contributed to the jet production.
More recently, \citet{chi08b} studied the properties of a recurring
AR jet observed in SXR and EUV and found a rapid decrease of
magnetic flux before the production of an SXR jet. They thus
concluded that the recurrent jets were powered by magnetic
cancellation.

Some models have been developed to explain jets and their associated
surges~\citep{sch95,can96}. Jets produced by magnetic reconnection
is a well acceptable viewpoint~\citep{shi07,jia11}, and is supported
by more and more observational results. Usually, magnetic
reconnection between newly emerging and pre-existing flux is
considered in numerous magnetohydrodynamic numerical
simulations~\citep{shi92,yok95,yok96}, in which the hot X-ray jet
and the cool adjacent H$\alpha$ surge could be simultaneously
produced. According to~\citet{shi96}, there are three possible
acceleration mechanisms of jets in the magnetic reconnection
process: evaporation flow jets, magnetic twist jets and reconnection
jets. Generally, evaporation flow jets are accelerated by the gas
pressure gradient force, and has velocities of order of sound
velocity. Magnetic twisted and reconnection jets are accelerated by
Lorentz force, and have velocities of order of Alf$\grave{e}$n
speed.

On 2007 June 5, there were approximately 18 coronal jets
continuously spouting from the northwest edge of NOAA AR
10960~(S08E29). Three of them were simultaneously observed by
$\emph{Hinode}$ XRT and EIS, and $\emph{TRACE}$ or $\emph{Solar
Terrestrial}$ $\emph{Relations Observatory}$ ($\emph{STEREO}$)
Extreme UltraViolet Imager (EUVI). In this paper, we compared the
morphological and kinetic characteristics of the three jets in SXR
and EUV wavelengths, derived their spectral characteristics, and
examined the evolution of photospheric magnetic field at their base.
This paper is organized as follows. The observational instruments
and data analysis are described in Section 2. The main observational
results are listed in Section 3, and conclusions and discussions are
presented in Section 4.

\section{Instruments and data analysis}
\label{instruments}

\subsection{Instruments and data}

EIS~\citep{cul07} instrument on $\emph{Hinode}$ observes the solar
corona and upper transition region in two EUV narrow wavebands:
170-210 and 250-290 {\AA}, with a spatial resolution along the slit
of $1''$\,pixel$^{-1}$ and a spectral resolution of
0.0223\,{\AA}\,pixel$^{-1}$. There are two spectral slits ($1''$ and
$2''$) and two spectral imaging slots ($40''$ and $266''$) for EIS
to imaging the Sun. Solar images can be constructed by moving a
slit/slot west to east across an interesting area with a given
exposure time at each step. The spectrum of each point can be
acquired by using the slits, while the monochromatic images can be
obtained by using the slots. In the present work, $2''$ slit was
used to scan an area of 240\arcsec${\times}$240\arcsec with 5s
exposure times, giving a duration time of 13 minutes. Each spectrum
at each location contained 17 spectral windows, and some of them
blended with several other lines. Our study primarily concentrated
on 6 emission lines, whose maximum ionization temperatures ranged
from 0.05 to 2.0 MK. Table 1 lists these emission lines (marked by
asterisk in table 1) included in this study and their maximum
ionization temperatures. There were three recurrent jets luckily
observed by four EIS raster sequences, at 01:51\,UT, 04:16\,UT,
10:11\,UT, and 10:24\,UT, respectively, while we only analyzed the
first three sequences. The EIS raster scanned each recurring jet in
a short time (within 3 minutes). Though propagating directions of
these jets were just opposite to the slit scanning orientation, we
considered that the final results could not be seriously affected
due to short scanning times.

$\emph{Hinode}$ XRT~\citep{gol07} is a grazing-incidence X-ray
telescope. It is equipped with nine filters on two filter wheels
located in front of the CCD detector, sensitive to plasmas with
temperature ranging from 1-30 MK. For the current study, SXR images
were provided by one single X-ray filter (Ti/Poly). The cadence of
these images was about twenty seconds, and the resolution was
$1''$\,pixel$^{-1}$. The field of view was $512''$${\times}$$512''$.

EUVI of the Sun Earth Connection Coronal and Heliospheric
Investigation~\citep{how08} instrument on board $\emph{STEREO}$
provides full-disk EUV images. The $\emph{STEREO}$ consists of two
space-based observatories, which orbit the sun ahead
($\emph{STEREO}$ $\emph{A}$) and behind ($\emph{STEREO}$ $\emph{B}$)
the Earth, respectively. The EUVI images are taken in four spectral
bands centered on Fe {\sc ix/x} (171\,{\AA}), Fe {\sc xii}
(195\,{\AA}), Fe {\sc xv} (284\,{\AA}), and He {\sc ii}
(304\,{\AA}). For the present work, the $\emph{STEREO}$ $\emph{A}$
EUVI 171\,{\AA} images were used. These images had a pixel size of
$1.59''$ and an alternative cadence of 2 and 3 minutes. Some 304,
195 and 284\,{\AA} images were also employed in our work, which had
the same resolution with the 171\,{\AA} images.

$\emph{TRACE}$~\citep{han99} provided 171\,{\AA} images with a 1
minute cadence and a $0.5''$\,pixel$^{-1}$ resolution for our study.
We also examined several $\emph{TRACE}$ white light (WL) and
1600\,{\AA} images, and $\emph{SOHO}$ Michelson Doppler Imager
(MDI)~\citep{sche95} full-disk 96-minute longitudinal magnetograms.

\subsection{Data reduction and Gaussion fitting}

\begin{table}[h!!!]

\small
\centering

\begin{minipage}[]{85mm}
\caption[]{ Hinode EIS Spectral Lines Analyzed in This
Study}\label{Table 1}\end{minipage}
 \tabcolsep 6mm
 \begin{tabular}{lcc}
  \hline\noalign{\smallskip}
Ion &  Wavelength({\AA}) & log$T_{max}$(K)\\
  \hline\noalign{\smallskip}
He{\sc ii}*   & 256.320 & 4.7 \\
Fe{\sc viii}* & 185.210 & 5.6 \\
Fe{\sc x}*    & 184.540 & 6.0 \\
Fe{\sc xii}   & 186.850+186.890 & 6.1 \\
Fe{\sc xii}*  & 195.120 & 6.1 \\
Fe{\sc xiii}* & 202.040 & 6.2 \\
Fe{\sc xv}*   & 284.160 & 6.3 \\
  \noalign{\smallskip}\hline
\end{tabular}
\end{table}

Firstly, we calibrated the raw EIS data by using the standard
processing routine $\emph{eis\_{}prep}$ in the SolarSoftWare (SSW)
packages~\citep{fre98}. This routine corrected the data for cosmic
ray hits, hot pixels, detector bias, and dark current. After this
correction, the absolute intensities in
ergs~cm$^{-2}$~s$^{-1}$~sr$^{-1}$~{\AA}$^{-1}$ were obtained. In
order to determine the spectral intensity, spectral line width and
centroid of each pixel, we approximately fit the Fe{\sc
viii}\,185.210, Fe{\sc x}\,184.540, Fe{\sc xii}\,195.120, Fe{\sc
xiii}\,202.040, and Fe{\sc xv}\,284.160\,{\AA} lines with a Gaussian
profile. Thereinto, the Fe{\sc xii}\,195.120 line was blended with a
weak Fe{\sc xii}\,195.180 line~\citep{li09}. This blend could not
affect our results significantly, so we fit the Fe{\sc xii}\,195.120
line by a single Gaussian. The Fe{\sc xii}\,186.880 line was only
used to estimate the electron density. Though it was a blend of
Fe{\sc xii}\,186.890 and Fe{\sc xii}\,186.850~\citep{you07a}, we
roughly fit it by a single Gaussian. The He{\sc ii} 256.320 line was
very complex, and it was blended with Si{\sc x}\,256.370, Fe{\sc
xiii}\,256.420, and Fe{\sc xii}\,256.410 lines~\citep{you07a}, which
could not simply be fit with a Gaussian profile. We adopted two
Gaussian fit to reduce the blending effect. In the fitting process,
we assumed that the He{\sc ii} line and the Si{\sc x} line had the
same width~\citep{lan09}, and the intensity ratio of the two lines
was 4~:~1, which was the minimum ratio estimated by \citet{you07a}.
The fitting results revealed that the two-Gaussian fit method was
feasible, and it really reduced the chi-square value.

To get believable spectral line centroids, we should note two
effects. One effect was slit tilt, which is caused by the tilt of
the EIS slits relative to the detector orientation, resulting in
line profiles at the bottom of the EIS slit being blue-shifted
relative to profiles at the top of the slit. The other effect was an
orbital variation of the spectral line position due to the
temperature variation, leading to the sinusoidal variation of
spectral line centroids. We eliminated the former effect by
performing the $\emph{eis\_{}tilt\_{}correction}$ routine. The
latter effect was removed via the $\emph{eis\_{}orbit\_{}spline}$
routine. This routine averaged the centroids along the quieter
region (bottom rows of 0-50) of each slit and got the time variation
of the averaged centroid. Then, a spline was fitted to this
variation, and the fit, as an orbit variation, was subtracted from
the centroids. This processing method dealing with the orbit effect
was also mentioned by~\citet{chi08a}. There was not an absolute
wavelength calibration in EIS. To compute the relative Doppler
velocities, we assumed that the reference wavelength is equal to the
average center of a quiet region, and the reliable Dopplergrams were
finally established. We calculated the electron density of the jets
by using the line intensity ratio of Fe {\sc
xii}~($\lambda$(186.850+186.890)/$\lambda$195.120) density-sensitive
pair. The non-thermal velocity was measured by removing the
instrumental width and the thermal width, and it can be calculated
by using the $\emph{eis\_{}width2velocity}$ routine.

\subsection{Data alignment}

We first corrected the pixel offsets between the long
(170-210\,{\AA}) and the short (250-290\,{\AA}) EIS wavelengths
images. By comparison, we shifted the short wavelengths images
$17''$ ($2''$) southward (westward) relative to the long wavelengths
images. The offsets were mainly owning to the instrumental offset
between the images taken in the two EIS CCDs~\citep{you07b}, and had
the same values as that given by \citet{chi08a}. We then co-aligned
the data from different instruments by comparing the solar features,
such as sunspots or bright patches. In order to co-align
$\emph{TRACE}$ images and MDI magnetograms, we chose a
$\emph{TRACE}$ WL image and an MDI continuum image at the closest
time and shifted the WL image in the solar X and Y directions,
respectively, until they were identical. With the same method, we
also co-aligned the 171\,{\AA} $\emph{TRACE}$ and EUVI images. For
EUVI and XRT images, we contrasted the simultaneous EUVI 284\,{\AA}
and XRT SXR images. We co-aligned the EUVI and EIS data by carefully
comparing the EUVI and EIS images at 195\,{\AA}.

\section{Observational results}
\label{sect:results}

\begin{figure}[!ht]
\centering
\includegraphics[width=10.0cm, angle=0]{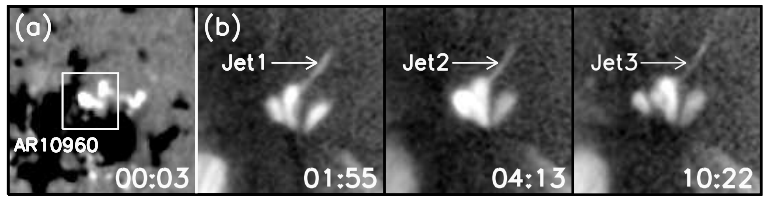}
\begin{minipage}[]{120mm}
\caption{General appearance of the jet region in $\emph{SOHO}$ MDI
magnetogram (a) and the three recurring jets in $\emph{Hinode}$ XRT
SXR images (b). The field of view (FOV) is
135\arcsec${\times}$135\arcsec.}
\end{minipage}
   \label{Fig1}
   \end{figure}

On 2007 June 5, the NOAA AR 10960 (S08E29) held a very complex
configuration of $\beta$$\gamma$$\delta$-type and was gradually
decaying. It is found that there were many coronal jets ejecting
from the north of its westernmost sunspot. Here, we mainly examined
three of them, which were simultaneously observed by EUV and SXR
wavelengths. Figure 1 shows the general appearance of the three
recurring jets on SXR images and their location on MDI magnetogram.
The NOAA AR 10960 was located at the lower left corner of these
images. It is noted that the three jets ejected with bright, nearly
collimated morphological structures from the same region, which
showed up as point like features on SXR images. Such kind of
footpoint was classified to XBP type by \citet{shim96}. They also
pointed out that jets from XBPs in ARs tended to appear at the
western edge of preceding sunspots, while our three jets appeared at
the north edge of preceding sunspot. From the magnetogram, we found
that these recurring jets located on a mixed polarities region
(marked by white box), which was consistent with the result from
\citet{shim98} that the XBP-like structures at the jet bases were
generally rooted in the mixed or satellite magnetic polarities
region. In the following section, we demonstrate that magnetic
emergences and cancellations take place in this region.

\subsection{EUV and SXR jets}

\begin{figure}[!ht]
\centering
\includegraphics[width=8.0cm, angle=0]{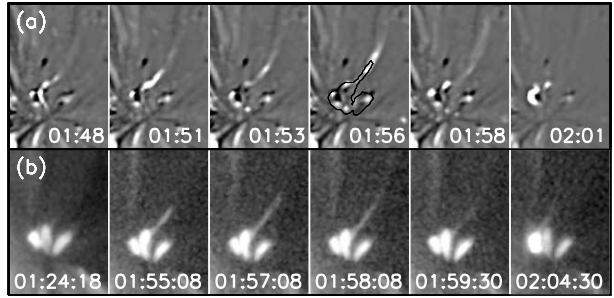}
\begin{minipage}[]{100mm}
\caption{$\emph{STEREO}$ EUVI 171\,{\AA} difference (a) and
$\emph{Hinode}$ XRT SXR direct (b) images showing the evolution of
the first jet. In a, all of the images are subtracted by the 00:06
UT image. The black contour plots the soft X-ray jet at 01:57:08 UT.
The FOV is 110\arcsec${\times}$160\arcsec.}
\end{minipage}
   \label{Fig2}
   \end{figure}

In order to explore morphological and kinetic relationships between
EUV and SXR jets, we present evolutions of the three EUV and SXR
jets in figure 2-4. Figure 2 shows evolutions of the first jet. In
this event, $\emph{STEREO}$ EUVI 171 {\AA} difference images were
used to show EUV jet more clearly. In EUV wavelength (Fig. 2a), we
can see a bright, linear jet structure ejected from a bright
footpoint. It stretched to the northwest with a converging shape.
This jet lasted for about 13 minutes (01:48-02:01~UT), and got the
maximum length of 8.1${\times}$10$^{4}$ km at 01:58~UT. We estimated
its velocity to be about 135 km s$^{-1}$. At the jet base, we
observed a very weak brightening, i.e., a small flare. It was a
common characteristic of jets and was called subflare or microflare.
We note that SXR observations (Fig. 2b) also revealed a bright jet
structure with a converging shape. Unfortunately, the XRT
observation missed the initial stage of the event due to a data gap
(01:24:18-01:55:08~UT). We found that SXR jet also reached the
maximum length at 01:58~UT, and its maximum length was about
8.3${\times}$10$^{4}$ km. If we assumed that the SXR and EUV jets
had the same start time, then the velocity of the SXR jet was about
138 km s$^{-1}$. In addition, the SXR jet at 01:57:08~UT was
superposed on the EUV image at 01:56~UT. It is clear that the SXR
and EUV jets had the same locations and directions.

\begin{figure}[!ht]
\centering
\includegraphics[width=8.0cm, angle=0]{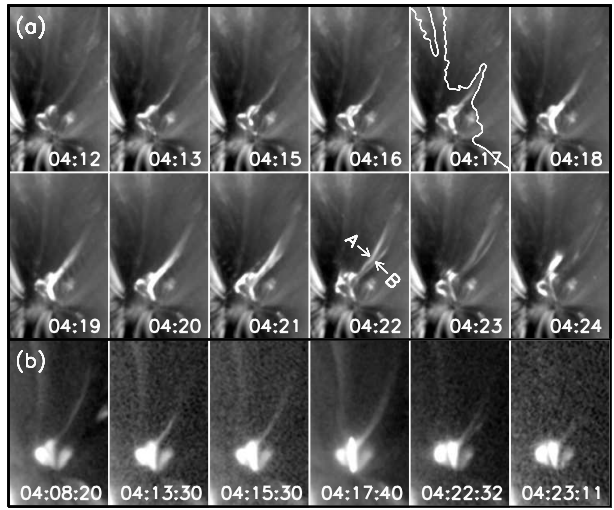}
\begin{minipage}[]{100mm}
\caption{$\emph{TRACE}$ 171\,{\AA} (a) and $\emph{Hinode}$ XRT SXR
(b) direct images showing the evolution of the second jet. The white
contour plots the soft X-ray jet at 04:17:40~UT. The FOV is
126\arcsec${\times}$200\arcsec.}
\end{minipage}
   \label{Fig3}
   \end{figure}

Evolutions of the second jet were shown in figure 3. In this case,
$\emph{TRACE}$ 171\,{\AA} images were used. This event was poorly
observed by $\emph{Hinode}$ XRT SXR observations, and lost data from
04:08:20 to 04:11:13~UT and from 04:17:40 to 04:20:33~UT. Unlike the
first jet, its accompanied small flare appeared to be more glaring
and its structure seemed to be more complicated. In figure 3a, the
EUV jet looked like a single thread before 04:21~UT, while it
divided into two threads (marked A and B) after 04:21~UT. Such a
bifurcation structure might result from the jet rotation, but we
could not identify such rotation from these EUV images. It is noted
that such bifurcation could also be seen clearly on the SXR image at
04:22~UT. The EUV jet started at 04:13~UT, reached to the maximum
length of 8.7${\times}$10$^{4}$ km at 04:21~UT and completely
disappear at 04:24~UT. Its velocity was calculated to be 145 km
s$^{-1}$. However, due to data gaps, the corresponding SXR jet lost
its start time and the time when it elongated to the maximum length.
Therefore, we could not give the size and velocity of the SXR jet.
We also superposed the SXR jet at 04:17:40~UT on the EUV image at
04:17~UT. As a result, we found that the SXR and EUV jets had the
same locations and directions.

\begin{figure}[!ht]
\centering
\includegraphics[width=8.0cm, angle=0]{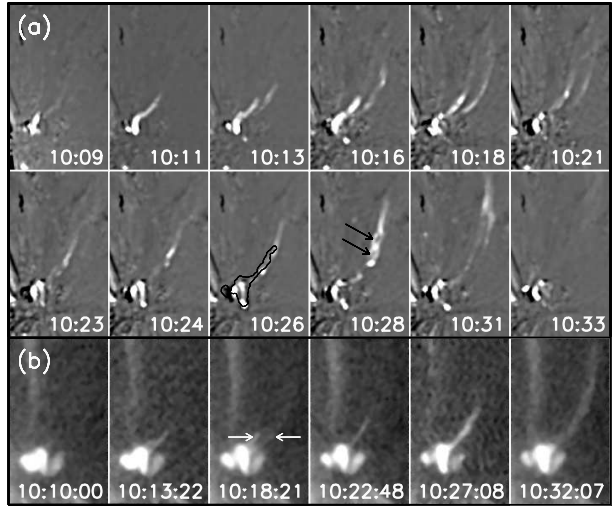}
\begin{minipage}[]{100mm}
\caption{$\emph{STEREO}$ EUVI 171\,{\AA} difference (a) and
$\emph{Hinode}$ XRT SXR direct (b) images showing the evolution of
the third jet. In a, all of the images are subtracted by the 10:01
UT image. The black contour plot the soft X-ray jet at 10:27:08 UT.
The FOV is 120\arcsec${\times}$200\arcsec.}
\end{minipage}
   \label{Fig4}
   \end{figure}

Different from the above two events, the third case contained two
jets. In figure 4a, we found that a bright elongated jet first
appeared at 10:11~UT. At around 10:12~UT, another jet below it
ejected from the jet base. Then, they propagated outward together.
At 10:18~UT, the second jet reached to the maximum length of
9.0${\times}$10$^{4}$ km, then quickly faded away and completely
disappeared at about 10:23~UT. We estimated its velocity to be about
201 km s$^{-1}$. Nevertheless, the first jet got its maximum length
of 1.3 ${\times}$ 10$^{5}$ km at 10:31~UT. The velocity of the first
jet was about 140 km s$^{-1}$. An interesting finding is that the
jet at 10:28~UT had two knots (marked by two white arrows), which
might indicate that the EUV jet had a twisted structure. The
subsequent bifurcation was clearly seen at 10:31~UT. It completely
disappeared after 10:33~UT. We also found two jets (marked by two
white arrows in figure 4b) in SXR. However, the SXR observations
also had many data gaps (10:13:22-10:18:21~UT, 10:18:41-10:22:48~UT
and 10:27:08-10:32:07~UT.). Due to these data gaps, we could not
calculate the maximum length and velocity of the second jet. If we
assumed that the first jet got the maximum lengths of 1.4 ${\times}$
10$^{5}$ km at 10:32~UT, then its velocity was estimated to be 144
km s$^{-1}$. We could not identify the twisted structure and
subsequent bifurcation on SXR images. Like the above two events, We
superposed the SXR jet at 10:27~UT on the EUV image at 10:26~UT. It
is found that the SXR and EUV jets had the same locations and
directions.

\subsection{Spectral characteristics}

\begin{figure}[!h]
\centering
\includegraphics[width=7.0cm, angle=0]{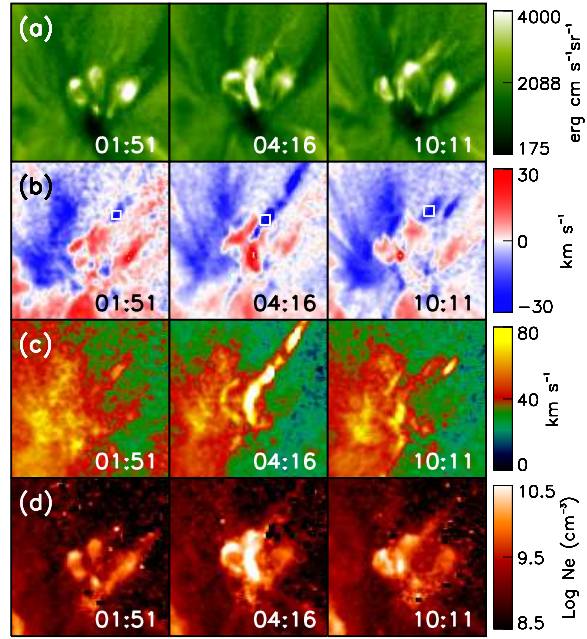}
\begin{minipage}[]{95mm}
\caption{$\emph{Hinode}$ EIS Fe{\sc xii} $\lambda$195\,{\AA}
intensity (a), Doppler velocity (b), non-thermal velocity (c), and
density (d) maps showing the three recurring jets. The FOV is
110\arcsec${\times}$110\arcsec.}
\end{minipage}
   \label{Fig5}
   \end{figure}

The three jets mentioned above were also observed by $\emph{Hinode}$
EIS, and their spectral characteristics were obtained. Figure 5
shows Fe{\sc xii} $\lambda$195\,{\AA} intensity, Doppler velocity,
non-thermal velocity, and density maps of the three recurring jets.
The Fe{\sc xii} $\lambda$195\,{\AA} line has the maximum ionization
temperature, i.e., the formation temperature, of 1.3 MK. From
intensity maps, three faint elongated jet structures were
identified. In the Doppler velocity maps, we note blue-shifted
features stretched above red-shifted features at footpoint regions.
Clearly, the blue-shifted features were corresponding to the three
jets. Maximum blue-shifted velocities and red-shifted velocities for
the three jets were calculated. Maximum blue-shifted velocities were
25, 121 and 43 km s$^{-1}$ and maximum red-shifted velocities were
11, 38 and 35 km s$^{-1}$. We also found non-thermal velocities for
the three jets and their footpoint regions. Maximum non-thermal
velocities associated with the three jets were 98, 181 and 98 km
s$^{-1}$, and maximum non-thermal velocities associated with
footpoint regions were 90, 130 and 90 km s$^{-1}$. Moreover, we also
estimated the maximum densities for the three jets, and the values
were 2.6${\times}$ 10$^{10}$, 3.4${\times}$ 10$^{10}$ and
6.6${\times}$ 10$^{9}$~cm$^{-3}$. In addition, the three jets could
be simultaneously observed by the six spectral lines listed in Table
1. Therefore, we considered that the three jets had temperatures
ranged from 0.05 to 2.0 MK.

\begin{figure}[!ht]
\centering
\includegraphics[width=7.0cm, angle=0]{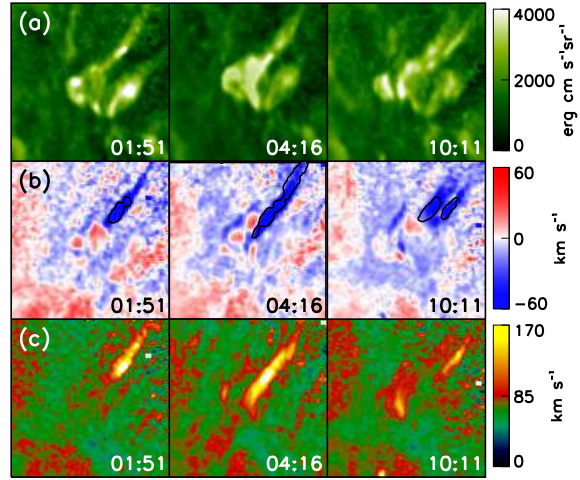}
\begin{minipage}[]{95mm}
\caption{$\emph{Hinode}$ EIS He{\sc ii} $\lambda$256\,{\AA}
intensity (a), Doppler velocity (b), and non-thermal velocity (c)
maps showing the three recurring jets. The three black contours
outline the three jets in Fe{\sc xii} $\lambda$195 {\AA} velocity
maps. The FOV is 110\arcsec${\times}$110\arcsec.}
\end{minipage}
   \label{Fig6}
   \end{figure}

We present the spectral characteristics of the three recurring jets
obtained by He{\sc ii} $\lambda$256\,{\AA} line (figure 6) to
compare with those obtained by Fe{\sc xii} $\lambda$195\,{\AA} line.
The He{\sc ii} $\lambda$256\,{\AA} line formed at a temperature of
0.05 MK. The three jets in the intensity maps given by He{\sc ii}
$\lambda$256\,{\AA} line appeared to be more powerful than that
given by Fe{\sc xii} $\lambda$195\,{\AA} line. We found blue-shifted
features related to the three jets and red-shifted features
associated with their footpoint regions. Their maximum blue-shifted
velocities were 229, 232, and 115 km s$^{-1}$, and maximum
red-shifted velocities were 21, 28, and 35 km s$^{-1}$. Maximum
non-thermal velocities associated with the three jets were 399, 336
and 196 km s$^{-1}$, while maximum non-thermal velocities associated
with the footpoint regions were 113, 263 and 155 km s$^{-1}$.
Apparently, maximum blue-shifted and non-thermal velocities
associated with the three jets calculated in He{\sc ii}
$\lambda$256\,{\AA} line were higher than those calculated in Fe{\sc
xii} $\lambda$195\,{\AA}.

\begin{figure}[!ht]
\centering
\includegraphics[width=8.0cm, angle=0]{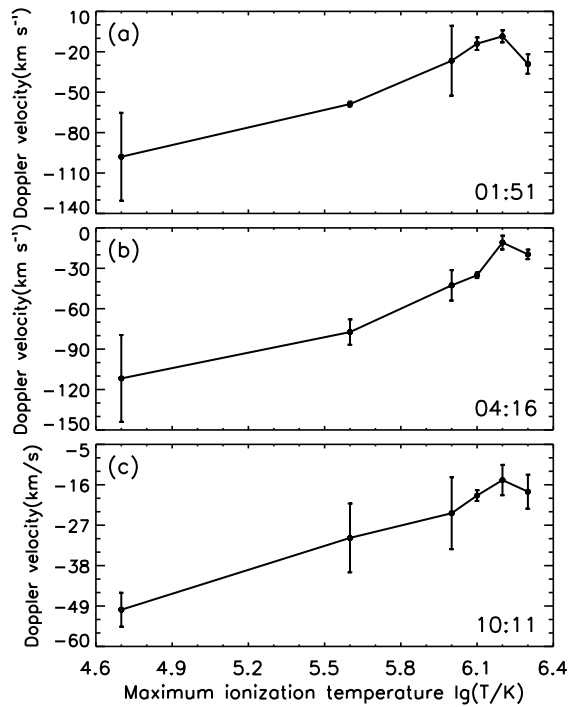}
\begin{minipage}[]{100mm}
\caption{The maximum ionization temperature dependence of the
Doppler velocity in the three jets averaged over the white boxes
marked in Fig.~5. The error bars represent the corresponding fitting
errors.}
\end{minipage}
   \label{Fig7}
   \end{figure}

Using six emission line listed in table 1, we investigated the
relations of averaged Doppler velocities and maximum ionization
temperatures for the three jets. For each jet, averaged Doppler
velocities were calculated in the white box in Figure 5 and these
relations for the three jets were shown in figure 7. We found that
the three jets had a common tendency: averaged Doppler velocities
decreased with the increase of the maximum ionization temperature.
Meanwhile, we superposed blue-shifted feature corresponding to the
three jets in Fe{\sc xii} $\lambda$195 {\AA} Doppler velocity maps
on He{\sc xii} $\lambda$256 {\AA} Doppler velocity maps (see figure
6b) and found that the blue-shifted area also decreased with the
increase of the maximum ionization, which was consistent with the
result given by \citet{kim07}. According to these observational
results, we infer that magnetic reconnection associated with these
recurring jets take place at the lower transition region or
chromosphere, i.e., low temperature region.

\subsection{Evolutions of longitudinal magnetic fields}

\begin{figure}[!ht]
\centering
\includegraphics[width=10.0cm, angle=0]{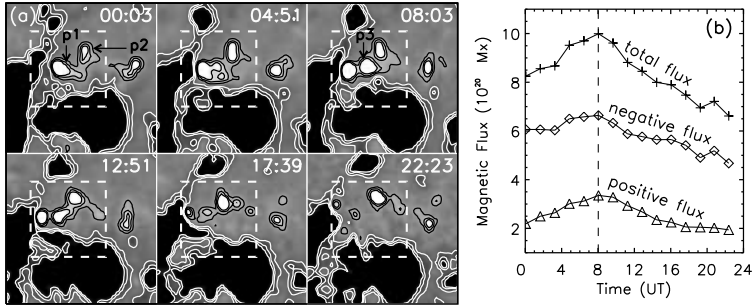}
\begin{minipage}[]{120mm}
\caption{(a) Time sequences of $\emph{SOHO}$ MDI magnetograms
superposed by iso-gauss contours with levels $\pm$50, $\pm$100 and
$\pm$150G. White boxes indicate the base of these recurring jets.
(b) Evolutions of positive, negative and total magnetic fluxes in
white boxes of this figure. The FOV is 80\arcsec${\times}$80\arcsec.
The dashed vertical bar represent the time boundary of the increase
and decrease of the magnetic flux.}
\end{minipage}
   \label{Fig8}
   \end{figure}

The magnetic field evolution at the base of these recurring jets
(marked by white boxes) was shown in figure 8a. The positive and
negative polarities were represented by white and dark colors,
respectively. For clearly exhibiting the magnetic field evolution,
we superposed iso-gauss contours with levels $\pm$50, $\pm$100 and
$\pm$150G on these magnetograms. We note two emerging magnetic
features, "p1" and "p2", at the jet base in the 00:03~UT frame.
Before 08:03~UT, their area continuously increased, indicating that
new magnetic flux emerged into this region. Meanwhile, these
emerging magnetic features moved southeastward, then squeezed and
cancelled with their surrounding negative magnetic flux. At
08:03~UT, the positive magnetic flux got the maximum area and
another emerging magnetic feature, "p3", between p1 and p2 were
identified, while the negative magnetic flux east of them clearly
decreased. After 08:03~UT, the positive magnetic flux and their
surrounding negative magnetic flux apparently decreased. At
22:23~UT, p1 and p3 almost disappeared completely.

We also calculated the positive, negative and total magnetic fluxes
in white boxes of Figure 8a and showed their evolutions in figure
8b. It is obvious that the opposite magnetic flux increased
asymmetrically before 08:03~UT, with positive magnetic flux
increased largely whereas the negative magnetic flux increased
slightly. According to our calculation, average flux-gain rates of
the positive flux and the negative flux were 4.0 ${\times}$
10$^{15}$ Mx s$^{-1}$ and 2.0 ${\times}$ 10$^{15}$ Mx s$^{-1}$,
respectively. After 08:03~UT, the decrease of the opposite magnetic
flux also show an asymmetry, and flux-loss rates of the negative
magnetic flux and the positive magnetic flux were 3.8 ${\times}$
10$^{15}$ Mx s$^{-1}$ and 2.7 ${\times}$ 10$^{15}$ Mx s$^{-1}$,
respectively. For the asymmetry before 08:03~UT, it might be the
result that the negative polarity of the emerging bipole was out of
the field marked by white boxes. The asymmetry after 08:03~UT can be
interpreted as part of negative magnetic flux moving out the white
boxes. As a result, the total magnetic flux had a averaged flux-gain
rate, 6.1 ${\times}$ 10$^{15}$ Mx s$^{-1}$, and a averaged flux-loss
rate of 6.6 ${\times}$ 10$^{15}$ Mx s$^{-1}$. According to
\citet{shim98}, the difference between the observed increase or
decrease in the flux of the jet-producing region was resulted from
the difference in the rates of the magnetic flux emergence and
photospheric reconnection. Furthermore, we found that these
recurring jets occurred not only when total magnetic flux was
increasing, but also when it was decreasing, which was similar to
observations of \citet{shim98}.

\section{Conclusions and discussions}
\label{sect:conclusions}

In this paper, we compared three EUV and SXR jets and analyzed their
spectral properties, and the main observational results are
summarized as follows: (1) These recurring jets were located on a
mixed polarities region near a sunspot, and clear magnetic flux
emergence and cancellation occurred at their location. (2) These EUV
and SXR jets ejected from the same site, and had similar directions,
sizes and velocities. The third EUV jet showed an apparent helical
structure, but this structure could not be resolved on SXR images.
(3) These three jets had temperatures from 0.05 to 2.0 MK and
maximum electron densities from 6.6$\times$10$^{9}$ cm$^{-3}$ to 3.4
$\times$10$^{10}$ cm$^{-3}$. (4) Elongated blue-shifted features
associated with jets and red-shifted features at the jet bases were
simultaneously observed in Fe{\sc xii} $\lambda$195 and He{\sc ii}
$\lambda$256 lines. (5) In Fe{\sc xii} $\lambda$195, the three jets
had maximum Doppler velocities from 25 km s$^{-1}$ to 121 km
s$^{-1}$ and maximum non-thermal velocities from 98 km s$^{-1}$ to
181 km s$^{-1}$; the jet bases had maximum Doppler velocities from
11 to 35 km s $^{-1}$ and the maximum non-thermal velocity from 90
to 130 km s$^{-1}$. (6) In He{\sc ii} $\lambda$256, maximum Doppler
velocities and maximum non-thermal velocities of the three jets were
from 115 to 232 km s$^{-1}$ and from 196 to 399 km s$^{-1}$,
respectively; maximum Doppler velocities and maximum non-thermal
velocities at the jet bases were from 21 to 35 km s $^{-1}$ and from
113 to 263 km s$^{-1}$, respectively. (7) averaged Doppler
velocities associated with jets decreased with the increase of
maximum ionization temperatures.

We compared EUV and SXR jets in an AR, and found that they had
similar sizes and velocities. These observational results were
consistent with \citet{kim07}, who discovered that EUV and SXR jets
had similar characteristics in terms of projected speed, size and
lifetime. However, these characteristics were different from
\citet{cha99}. In their work, the EUV jet had a typical size of
4000-10000km, a transverse velocity of 50-100km s$^{-1}$, and a
lifetime of 2-4 minutes, smaller and shorter lived than X-ray jets.
According to \citet{kim07}, these differences were probably caused
by the different locations of the reconnection points. Magnetic
reconnections related to our investigated recurring jets might take
place at more higher solar atmosphere than those studied by
\citet{kim07}. In addition, one of the three EUV jets implied a
helical structure, while the corresponding SXR jet did not. The
differences in morphology between EUV and SXR jets were ever
reported by \citet{ale99}, \citet{kim07} and \citet{chi08b}.
\citet{ale99} pointed out that EUV jets had twisted structures,
\citet{kim07} demonstrated multi-structures in EUV jets, while
\citet{chi08b} observed expanding loops associated with X-ray jets
instead of EUV jets.

For the three jets, their corresponding Doppler velocity maps from
EIS revealed apparent blue-shifted features elongated from
red-shifted bright points. We interpreted them as jets because all
of them were transient. Their maximum Doppler velocities were
calculated and it is found that all the derived values were lower
than that given by \citet{chi08a}, who showed that the upflow
velocities associated with an active region jet exceeded 150 km
s$^{-1}$. In addition, the maximum Doppler velocity obtained by
He{\sc ii} 256.320 line could reach to 232 km s$^{-1}$, which was
much larger than 62 km s$^{-1}$ estimated by \citet{kim07}.
Moreover, We examined the relationship between the averaged Doppler
velocity and the maximum ionized temperature near the jet base, and
noted that the averaged Doppler velocity decreased with the increase
of the maximum ionized temperature. According to this observation,
we assumed that magnetic reconnections associated with these three
jets might occur in the low-temperature region, i.e., the higher
chromosphere or the lower transition region. \citet{kim07}
simultaneously found blue-shifted and red-shifted motions at the jet
base. Different from them, we only observed red-shifted features at
the jet base, which were similar to the observations of
\citet{kam07} and \citet{chi08b}. Because evaporation flow was
considered to be an upflow at the coronal temperature, this
phenomenon was not consistent with the evaporation scenario
\citep{kam07}. We regarded this kind of red-shifted feature as
downflows in reconnected flares, downward reconnected loops, or
downward streams of bi-directional reconnection outflows.

In this work, obvious magnetic flux emergences and cancellations
induced these recurring jets. Such instances were ever reported by
\citet{cha99}, \citet{zha00}, \citet{liu04}, and \citet{jia07}.
Recently, \citet{arc10} proposed a 3D numerical simulation, in which
a toroidal flux tube emerged into the solar atmosphere and
interacted with a pre-existing field of an AR. In their simulation,
recurrent coronal jets were produced. They thought that the
emergence of new magnetic flux introduced a perturbation to the AR,
and caused reconnection between neighboring magnetic fields and the
release of the trapped energy in the form of jet-like emissions.
Moreover, they also pointed out that when the amount of emerging
flux was exhausted, the dynamic rise slowed down and reached an
equilibrium. In our case, magnetic emergences first dominated the
jet base, the amount of them reached to the maximum at 08:03~UT.
Then they began to decay, and their associated jet activities
decreased. At 22:23~UT, emerging magnetic features p1 and p3 almost
completely disappeared. After 22:23~UT, we could not observed one
clear jet activity. These observations strongly support the
simulation results given by \citet{arc10}.

\normalem
\begin{acknowledgements}
The authors are indebted to the $\emph{Hinode}$/EIS,
$\emph{Hinode}$/XRT, $\emph{STEREO}$/EUVI, and $\emph{SOHO}$/MDI
teams for providing the wonderful data. $\emph{Hinode}$ is a
Japanese mission developed and launched by ISAS/JAXA, with NAOJ as
domestic partner and NASA and STFC (UK) as international partners.
It is operated by these agencies in co-operation with ESA and NSC
(Norway). We would like to thank the referee for constructive
comments, thank professor Peter R. Young for providing methods to
deal with blend effects of spectral lines, and thank Zhongquan Qu,
Kejun Li and Dengkai Jiang for valuable and helpful discussions.
This work is supported by the 973 Program (2011CB811403), by the
Natural Science Foundation of China under grants 10973038 and
40636031, and by the Scientific Application Foundation of Yunnan
Province under grants 2007A112M and 2007A115M.
\end{acknowledgements}

\label{lastpage}

\end{document}